Modelling of light driven $CO_2$ concentration gradient and photosynthetic carbon assimilation flux distribution at the chloroplast level

M. Jouravlev

POSTECH, San 31, Hyojadong, Namgu, Pohang 790-784, Korea

*e-mail: jouravl@rambler.ru

Abstract

The steady state of the two-substance model of light driven carbon turnover for the photosynthetic $CO_2$ assimilation rate is presented. The model is based on the nonlinear diffusion equation for a single chloroplast in the elliptical geometry by assuming light driven Ribulose-1,5-bisphosphate (RuBP) regeneration and $CO_2$ assimilation reaction of carboxilation coupled with the photosynthetic sink strength. The detailed analysis of 3 -dimensional $CO_2$ concentration and flux on the chloroplast level is made. It is shown that under intense light irradiation there exists a boundary layer of chloroplasts with a high value of $CO_2$ assimilation flux. The presented simplified model can be used for the calculations and experimental estimations of the $CO_2$ assimilation rate for environmental applications.

# 1. **Introduction**

In the present work we develop the theory for a new description and understanding of the limiting factors of plant photosynthesis and its components in order to develop quantitative analyses of the processes which can be used to predict how illumination conditions within the chloroplast affect carbon assimilation. The aim is to define the light dependent $CO_2$ diffusion in the chloroplast as limiting factors of carbon assimilation. These analyses are based on simplifications of the model light and dark reactions.

The combination of analysis of dark reaction and $CO_2$ distribution on the chloroplast level improves the understanding of the dynamics of photosynthetic regulation including the limiting factors or saturation. The potential factors limiting the rate of carbon assimilation in photosynthesis are the intensity distribution of light illumination and $CO_2$ diffusion in the volume of the chloroplast, which are both taken into account. The photosynthesis process in $C_3$-plants is realized by the following subsystems: fast subsystems and slow subsystems.

Fast subsystems are described by fast light reactions that take place on the membrane of thylakoids. Slow subsystems consist of dark chemical reactions of the Calvin cycle which includes binding of $CO_2$, chemical transport of $CO_2$ to sucrose, and regeneration of 3-phosphoglyceric acids (PGA) to RuBP. Photosynthesis of assimilates of carbon in $C_3$ plants takes place primarily in the chloroplast stroma, which contains many metabolites of the Calvin cycle and enzymes. Carbon assimilation is regulated by highly integrated mechanisms which allow the photosynthetic system to maintain its activity at rates appropriate to the demands of and changing conditions within the plant. All these processes are directly or indirectly complex functions of $CO_2$ concentration, metabolite concentrations and light intensity, which are highly variable and change often rapidly in complex ways.

The rate of $CO_2$ fixation is determined by the rate of turnover of the Calvin cycle metabolites, to supply the substrates and the conditions in the chloroplast. This article examines the mechanism of $CO_2$ assimilation and the associated process of $CO_2$ diffusion, and how the characteristics of different components in the light reactions interact to achieve such a highly regulated and effective system of assimilate production.

The main idea of the model involved light receptors with two states: the ground and excited state of the photosynthetic mechanism (Nitzan, 1973; Thornley, 1974). The receptors of the light receiving system are excited by the flow of photons with intensity $I$ to transform the receptors from the ground state to the excited state. The model consists of dynamical aspects of two-stage Thornley's model applying formal chemical kinetics with a known reaction rate constant and stoichiometry taking in to account the $CO_2$ diffusion process and hence time relaxation processes related to $CO_2$ transport. According to Thornley's model, the photon energy absorbed by the light receptors is included in an intermediate step of the Calvin cycle to the sucrose synthesis. The spatial distribution of $CO_2$ inside the chloroplast has to be considered.

## 2. The model

Photosynthesis in leaves is accomplished by a chain of chemical and photochemical reactions producing sugar. The main idea of this model is to describe the photosynthesis in the chloroplast as a chain of two reactions.

The structure of the chloroplasts is a membrane encompassing a system of thylakoids packed closely as granum – these granum are connected to each other by lamella. The granum and lamella are immersed in stroma. Light reactions take place in the membrane systems of chloroplasts. ATP and NADPH are synthesized by light energy. Chemical reactions of $CO_2$ binding take place in the stroma of the chloroplast.

The structure of the model is shown in Fig.1. Compared with the earlier version (Kaitala et al. 1982, Giersch, 2003.), two sub-stations of the Calvin cycle are considered, as well as $CO_2$ diffusion is included at the chloroplast level. Photosynthetic carbon metabolism is simplified drastically so that only turnover of carbon and inter-conversion of Calvin cycle intermediaries and of ATP ($S^*$-compound) and ADP (S-compound) are considered. The pools of Calvin cycle intermediaries are lumped together, with only the two species, RuBP (X-compound) and PGA (Y-compound), assumed to exist.

The absorption of light energy in the photosynthetic pigments causes photochemical events in which electrons are transferred along the series of molecules leading to the conversion of low energy compound S to high energy compound $S^*$. The reaction can be expressed as follows:

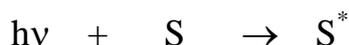

$$h\nu \;+\; S \;\rightarrow\; S^*$$

where $h\nu$ is the energy of light quanta. As provided by the well known Z-scheme of electron transport, the equation indicates that the light energy causes electron flow from the donor to acceptor with formation of high energy compound $S^*$ from S. This is described as an ordinary reaction but its rate of reaction is assumed to be a function of light intensity. The rate of the light driven reaction may be altered both by changing the light intensity and by light modulation of the concentration of S. The rate of the formation of $S^*$ is proportional to the product of the radiant flux density I and the concentration of S.

The X regeneration chain is simplified. It is assumed that Y (trioses) can be directly converted into X (pentoses), and hypothetical compound T provides the output of one carbon atom from the cycle. The following is a general equation for the regeneration reaction.

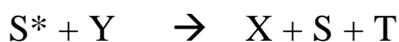

$$S^* + Y \;\rightarrow\; X + S + T$$

The $CO_2$ acceptor X is regenerated from Y. The role of the reaction is to regenerate X for further $CO_2$ assimilation. The chemical energy of the S* compound is assumed to be consumed in the course of a complete turnover of reduction cycle S* to S. The energy is expected to be required for the rearrangement of carbon atoms between Y and X, and synthesis of T compound (sugar). It is assumed that the turnover of one $CO_2$ molecule is included as a stoichiometric factor.

Carbon dioxide assimilation is a cyclic, autocatalytic process introduced by a chain of chemical reactions where X-substance produces Y- substance as follows:

$$X + CO_2 \rightarrow Y$$

This reaction emphasizes the carbon reduction and cyclic aspects of photosynthesis. The assimilation of $CO_2$ is described by carbamylation reaction. This reaction simplifies the complexity of the process.

In the carboxilation reaction catalysed by the enzyme RuBP carboxylase-axygenase (Rubisco), an acceptor molecule X, ribulose bisphosphate (RuBP) combines with $CO_2$, producing Y.

The stoichiometry of the reactions, as mentioned above is

$$X([5xC]) + CO2 (1xC) \rightarrow Y (2 [3xC])$$

$$Y (2[3xC]) \rightarrow X([5xC]) + T(1xC).$$

where X (RuBP) consists of 5 carbons atoms and 1 captured carbon atom are combined with Y (PGA) and T (TP). Turnover of X is stoichiometrically coupled to turnover of Y and vice versa. One carbon atom involved in the cycle is fixed in the kind of compound T.

The rate of cycle turnover in the steady state depends on the rates of $CO_2$ assimilation on enzyme activities, and availability of $CO_2$. This approach makes it possible to describe the light driven RuBP regeneration and $CO_2$ diffusion in chloroplasts. The conditions in the chloroplast which are necessary for the activation of RuBP regeneration depend on $CO_2$ distribution within the chloroplast. It is necessary to take into account the spatial distribution of $CO_2$ inside the chloroplast, as well as that the carbamylation occurs at small $CO_2$ concentrations.

The presence of RuBP produces binding of Rubisco to form inactive complexes such that the $CO_2$ assimilation efficiency is reduced. There are factors inhibiting $CO_2$ assimilation: light and diffusion processes inside the chloroplast. It can limit the rate of response of $CO_2$. This model replaces the Thornley's model and reproduces specific features of the photosynthesis process such as saturation at high light intensities and high $CO_2$ concentration in chloroplasts as well as sink strength nonlinear dependences on light intensity.

Other factors included are:

1. The rate of light-driven RuBP regeneration depends both on light intensity and rate of biochemical reaction of RuBP regeneration.

2. An obvious generalization of the biochemical reaction of the Calvin cycle is to assume that the rate of $CO_2$ fixation is proportional to the concentration of X, and the rate of RuBP regeneration is proportional to the light intensity.

3. The chloroplasts homogeneous structure consists of the granum and the lamellas; the thickness of each is omitted.

4. The concentration of NADP and $NADPH^+$ is assumed to be large at any moment of time, and the rate of conversion of NADP to $NADPH^+$ is large in comparison with the conversion rate of ADP to ATP, because two electrons are required for the generation of one $NADPH^+$. The rate of electron transport required to sustain the necessary $NADPH^+$ concentration is more than the rate of ATP production i.e. the production rate per 1 electron.

The units used throughout the model are molar ($M/m^3$) for concentrations and $(M/m^3)^{-1}s^{-1}$ for the reaction rates. A more complete discussion of the basic model design can be found in the Refs. (Kaitala, 1982; Giersch, 1986; Lushnikov et al., 1997; Pettersson et al., 1988.; Massunaga, et al., 2001.)

## 3. Model analysis

### 3.1 Basic equations

In order to analyze the chloroplast-level model we used the dynamical model of two biochemical processes: light interaction with the molecular species and two substances of reaction with $CO_2$ (Thornley, 1974, 1976; Giersch, 1986; Lushnikov, 1997). $CO_2$ carboxylation can be limited in two different ways depending on the $CO_2$ concentration and the irradiance intensity as well the RuBP regeneration-limited rate of net $CO_2$. It is supposed that the Rubisco activity-limited rate has a linear dependence on the rate of $CO_2$ carboxylation at low $CO_2$ concentrations and on high light intensity. Thus, the Rubisco activity-limited rate is included by implicit linearization in the constants of the model parameters (Lushnikov, 1997). According to the 3D optical model of photon transport in a leaf (Ustin et al., 2001), the main part of incident light energy (0.8 $I$ and less, where $I$ is the incident light intensity) is adsorbed by mesofill cells; nearly 100 percent of the rays are absorbed in leaf tissue i.e. within the initial 90 μm of the adaxial leaf surface and the gradient in light absorption of 0.3 and less within the leaf when illuminated at 675 nm, the photochemical reaction of excitation of harvesting light complexes has the linear dependence on $I$. Thus the photochemical reaction and the net $CO_2$ assimilation rate can therefore be expressed as the form of two differential equations which can be written as follows (Nitzan, 1973; Kaitala, 1982; Lushnikov, 1997):

$$\frac{\partial S^*}{\partial t} = k_I I (S - S^*) - k_Y Y S^* \qquad (1)$$

where $k_I$ is the photon absorption rate, $k_Y$ is the rate constant of RuBP regeneration, $S^*$ is the ATP concentration and $S$ is the ADP concentration in the chloroplast. According the reduced photosynthetic cycle shown in Fig.1, the kinetic equations for the concentration of intermediates are as follows:

$$\frac{\partial X}{\partial t} = -k_X X C(\vec{r}, t) + k_Y Y S^* \qquad (2)$$

$$\frac{\partial Y}{\partial t} = -k_Y Y S^* + k_X X C(\vec{r}, t) \qquad (3)$$

where $X$ is the RuBP concentration, $Y$ is the concentration of PGA, $k_X$ is the rate of carboxilation, $C(\vec{r}, t)$ is the time and spatial dependent distribution of $CO_2$ concentration, and $\vec{r}$ is the radius vector inside the volume of the chloroplast. Eqn.(1) contains two terms describing light-driven ATP generation and their consumption in dark reactions.

Eqn.(1) is a linear equation of first order for the dynamics of light activation of harvesting light complexes showing the rate of change in absorbing radiant intensity $I$ and turning it into so-called assimilatory power in the form of chemical energy. Here the same letters as for reactants stand for concentrations. $I$ is the light intensity (in principle, it is possible $I^n$). Let us say $n = 3$ or 4 if the process requires three or four light quanta. Eqn.(1) describes the change in time $t$ of the concentration of high-energy compound $S*$ and $S$ written with the following assumptions: (i) the light absorption and subsequent radiationless relaxation processes occur on a time scale much shorter than that for the $X \leftrightarrow Y$ reaction, and (ii) the radiationless relaxation processes are fast relative to the optical excitation and de-excitation $S + h\nu \rightarrow S^*$ rates and $S^* \rightarrow X$, so that the modification of the rate coefficient due to this last reaction may be disregarded. In addition to differential equations, the conservation equation for $S*$ and $S$ in the chloroplast has the form:

$$S^* + S = S_m \quad (4)$$

where $S_m = const$ is the net concentration of high energy metabolites: ADP and ATP. In the general case $S_m$ is a time-independent parameter of the model and it depends only on the light intensity $I$. Binding of $CO_2$ is the independent process of the photochemical cycle. Eqns. (2) and (3) described the change and uptake of $CO_2$ in time and RuBP regeneration from trioses. The conservation equation implies that the concentrations of $X$ and $Y$ remain constant for all time, so that:

$$\frac{\partial}{\partial t}(X + Y) = 0 \qquad (5)$$

thus

$$X + Y = X_0 \quad (6)$$

were $X_0$ is the total concentration of $X$ and $Y$ intermediates. The kinetics of the $CO_2$ consumption is described by the diffusion-reaction differential equation:

$$\frac{\partial C(\vec{r}, t)}{\partial t} - D\Delta C(\vec{r}, t) = -k_X X C(\vec{r}, t) \qquad (7)$$

here $\Delta$ is the Laplas operator, and $D$ is the $CO_2$ diffusion coefficient . Eqn. (7) is supplemented with the boundary condition:

$$C(\vec{r},t)\big|_{\vec{r}=\vec{r}_0} = c_0 \qquad (8)$$

where $c_0$ is the constant of concentration of $CO_2$ on the boundary of the chloroplast and can be derived from the initial condition of $CO_2$ concentration in the mesophyll (Cooke, 1967; Lushnikov, 1994), see Table 1.

The last term of equation (7) describes the sink strength in the form:

$$A = k_X X C(\vec{r},t) \qquad (9)$$

The sink strength $A$ was treated as continuous through the chloroplast volume. The dependence of the reaction constant on the concentration is displayed explicitly. Hence, our task is to find the kinetics of the metabolites $X$ and $Y$ and then the consumption or sink strength of $CO_2$. In the following section we adopt the steady-state approach for the stationary solution of the diffusion equation (7). The stationary solution of eqn. (7) is legitimate if any changes in external parameters such as $I$ and $CO_2$ concentration are much slower than all internal processes eqns. (2) and (3).

3.2 Steady state analysis

The time derivatives of metabolite concentrations $X$, $Y$, $S$ and $S*$ vanish and the result of equations (1)-(3), (7) for the steady state is

$$k_I I(S - S^*) = k_Y Y S^* \qquad (10)$$

$$-k_X X C + k_Y Y S^* = 0 \qquad (11)$$

$$D \Delta C = k_X X C \qquad (12)$$

The steady state metabolite concentrations are thus the solution $X$ and $Y$ of the equations (10)-(12)

$$X = \frac{k_Y S^*}{k_Y S^* + k_X C} X_0 \qquad (13)$$

$$Y = \frac{k_X C}{k_Y S^* + k_X C} X_0 \qquad (14)$$

with $\xi = k_X / k_Y$ as the internal parameter of the model. The physical interpretation of $\xi$ is the relation of the carboxilation rate to the RuBP-regeneration rate. Thus the formulas (13) and (14) have the form

$$X = \frac{S^*}{S^* + \xi C} X_0 \qquad (15)$$

$$Y = \frac{\xi C}{S^* + \xi C} X_0 \qquad (16)$$

The steady-state values of the rate coefficients of light driven RuBP-regeneration $k_Y$ is the internal parameter of the model. It can not be measured directly by an experimental set up, however the analytical form of $k_Y$ is required. With $\gamma = S / S*$ from eqns. (10)-(11) we have the intensity dependent RuBP-regeneration rate

$$k_Y = \frac{k_I I (\gamma - 1)}{X_0} \left( 1 + \frac{S*}{\xi C} \right) \qquad (17)$$

A straightforward analysis of the values of the concentrations of the intermediates of Calvin cycles (Kaitala, 1982; Hahn, 1984, 1987, 1991; Milstein, 1979; Pettersson, et al. 1988; Poolman, et al. 2000; Milstein, and Bremermann, 1979.) leads to the following necessary limiting conditions of concentration of the intermediates for the steady states $\xi C/S* \gg 1$ (a) and $\xi C/S* \ll 1$ (b), see Table 1.

In the case (a) the eqns. (15)-(17) and (7) take the form

$$k_Y = \frac{k_I I (\gamma - 1)}{X_0} \qquad (18)$$

$$X = \frac{\alpha' I}{\alpha' I + \beta' C} X_0 \qquad (19)$$

$$Y = \frac{\beta' C}{\alpha' I + \beta' C} X_0 \qquad (20)$$

where

$$\alpha' = k_I (\gamma - 1) S* \qquad (21)$$

and

$$\beta' = k_X X_0 \qquad (22)$$

The $CO_2$ diffusion equation (7) and boundary condition (8) take the nonlinear form

$$D\Delta C = \frac{\eta' I C}{S* + \xi C} \qquad (23)$$

$$C\big|_\Sigma = c_0 \qquad (24)$$

where

$$\eta' = k_I(\gamma - 1)\xi S* \qquad (25)$$

and $\Sigma$ is the surface of the chloroplast. Combining eqns. (9) and (23) we obtain the simple formulae for the sink strength

$$A = \frac{\eta' I C}{S* + \xi C} \qquad (26)$$

The solution of the nonlinear equation of $CO_2$ diffusion (23) of steady state is most often approached by utilizing the linearization methods which state that the condition for stability of a given solution of a nonlinear differential equation is identical to the solution derived from that equation linearized around the same steady solution (Sveshnikov, 1993). Defining the variable function: $W = C_0 - C$ we have linearized the system of eqns. (23), (24) around a given state solution $C_0$. We obtain

$$-D\Delta W = \frac{\eta' I(C_0 - W)}{S* + \xi(C_0 - W)} \qquad (27)$$

$$W\big|_{\Sigma} = 0$$

(28)

Using the linear part of Taylor's series for (27) we obtained

$$-D\Delta W = \frac{-\eta' I C_0}{S* + \xi C_0} + \frac{\eta' I S*}{(S* + \xi C_0)^2} W$$

(29)

$$W\big|_{\Sigma} = 0$$

(30)

We recover eqns. (29) and (30) by the inverse transform of variable $C = C_0 - W$ in the form which is equivalent to

$$D\Delta C = \frac{\eta' S* I}{(S* + \xi C_0)^2} C + \frac{\eta' \xi C_0^2}{(S* + \xi C_0)^2} I$$

(31)

$$C\big|_{\Sigma} = c_0$$

(32)

Thus the steady state of $CO_2$ distribution in the volume of the chloroplast is the solution of the inhomogeneous Helmholtz equation (31) with the Dirichlet boundary condition (32). For the light dependent sink strength in the linearized form we have

$$A = \frac{\eta' I S*}{(S* + \xi C_0)^2} C$$

(33)

For the effective diffusion coefficient we have

$$D_{eff} = \frac{(S* + \xi C_0)^2}{\eta' I S*} D \qquad (34)$$

Taking into account eqn. (34) and using the Einstein formula for the time of diffusion for the effective diffusion relaxation time we have

$$\tau_{eff} = \frac{L^2}{D} \frac{\eta' I S*}{(S* + \xi C_0)^2} \qquad (35)$$

It has the physical interpretation as an active layer of the chloroplast where the main part of $CO_2$ is consumed which shows typical light dependent behavior. In the opposite case (b) the eqns. (15)-(17) and (7) take the form

$$k_Y = \frac{k_I I (\gamma - 1)}{X_0} \frac{S*}{\xi C} \qquad (36)$$

$$X = \frac{\alpha' I}{\alpha' I + \beta' C^2} X_0 \qquad (37)$$

$$Y = \frac{\beta' C^2}{\alpha' I + \beta' C^2} X_0 \qquad (38)$$

where

$$\alpha' = k_I (\gamma - 1)(S*)^2 \qquad (39)$$

$$\beta' = k_X X_0 \xi \qquad (40)$$

The $CO_2$ diffusion equation (7) and boundary condition (8) takes the linear Laplace form

$$D\Delta C = \frac{\eta' I}{S^* + \xi C} \qquad (41)$$

where

$$\eta' = k_I(\gamma - 1)(S^*)^2 \qquad (42)$$

and after linearization the diffusion equation has the form

$$D\Delta C = \frac{\eta' I}{S *} \qquad (43)$$

$$C\big|_\Sigma = c_0 \qquad (44)$$

Using eqns. (9) and (37) for the linearized sink strength we have:

$$A = \frac{\eta' I}{S^*} \qquad (45)$$

We have thus shown that under conditions of low light irradiance and low ambient $CO_2$ concentration the sink strength has the linear dependence on $I$ and it approximates any well known experimental curves of photosynthesis.

## 4. Discussion

The model is based on the general assumption of the biology of leaf photosynthesis: (i) the assimilation of $CO_2$ has increased nonlinearly as a function of the light irradiance intensity, leveling off when photon flux density has reached the saturation point (1000 μ mol quanta m$^{-2}$ s$^{-1}$ ). The well defined $CO_2$ assimilation versus photon flux density is approximated by means of a hyperbola, (ii) the boundary $CO_2$ concentration at the saturation level of $CO_2$ exchange rate is the sub-stomatal $CO_2$ concentration.

The kinetics of the two substance model with the light dependent $CO_2$ diffusion was previously described to be confined to the carbon assimilation flux. The $CO_2$ flux was examined on the level of one chloroplast and a whole leaf. In the description of the model (eqn. (17)) the storage flux was examined and it was found to be distributed under the conditions of high irradiance and intermediate concentration of the turnover cycle. Metabolic analysis was used to quantify the effect of altered light reaction activity by determining the assimilation flux and sink strength of $CO_2$. At present, it is possible to propose the two cases of parameters (a) and (b) presented in Table 1 corresponding to both high and low values of the boundary $CO_2$ concentration and the rates of ADP and RuBP production of the $CO_2$ pathway in the steady state.

The reaction $CO_2$ assimilation rate in the steady state is saturated with respect to the internal parameter γ, the ratio of ADP and ATP concentration, although ATP is known to be present in the chloroplast as it is increased in high light conditions and consumed by the action of RuBP production. This reaction also lowers ATP concentration by introducing the RuBP regeneration reaction in the following manner: the only reaction of light absorption by chlorophyll molecules or the ATP production is the light reaction (coefficient $k_l I$ ) which acts as the effective chemical energy production. Therefore any increase in steady state flux through the RuBP regeneration must be accompanied by an equivalent increase in the $CO_2$ flux through this reaction, regardless of the mechanism which brings the increase about. However, ADP is also a product of this dark reaction, and so any increase in the ADP leak must also result in an equivalent increase in the production of PGA (see Fig.1) and the gradient of $CO_2$ concentration occurring in the chloroplast, effectively by-passing RuBP.     As the calculations were made under light saturation conditions, it is assumed that ATP activity is sufficient to lower the $CO_2$ concentration, so the $CO_2$ flux distribution mechanism in the chloroplast should be sought. Such a mechanism is the $CO_2$ consumption pathway responsible for the synthesis of PGA from RuBP which provides the nonuniform $CO_2$ flux distribution in the chloroplast volume. As a major effort would be required to include the $CO_2$ flux distribution into the simplified Calvin cycle, the likely effect of light dependent activity can be investigated at least to a first approximation.

The driving force for the $CO_2$ diffusion is the concentration gradient corresponding to the rate of the $CO_2$ assimilation.   There is the concentration difference between the surface boundary layer of the chloroplast and the sub-stomatal cavity. Due to the reaction of carboxilation or $CO_2$ consumption, the gradient of $CO_2$ concentration inside the boundary layer of the chloroplast is established.   It provides the flux of $CO_2$ from the sub-stomatal cavity to the chloroplast with the effective diffusion coefficient dependent on the intensity of light irradiance.

In order to conduct the numerical simulations, the appropriate values of the rate constants of biochemical and physical quantities have to be determined. Some of these constants are well defined and the overall model is not sensitive to others; the determination of the relation of the rate coefficients is defined by eqns. (17), (18), (36). Table 1 presents a set of values that are included in the model. At present, the rate of light irradiance dependent ATP production $k_1 \cdot I$  and the RuBP regeneration rate $k_Y$ can be estimated both from the measurements and computer simulation of the Calvin cycle (Kaitala, 1982; Lushnikov, 1997; Milstein and Bremermann, 1979). The parameters of eqns. (1)-(3)  $k_Y$ and $k_X$ may be determined from the estimates in the measured values of $A_J$ and $I$. Employing equations (17), (33) and (45), we calculate X, Y e.t.c. using the constant value of intermediates of the Calvin cycle (Lushnikov, 1997; Milstein and Bremermann, 1979).

The geometric characteristic of the chloroplast are chosen as an ellipse. In order to solve eqns. (31) and (32), finite difference methods were employed as the iterative process for elliptical symmetric shape of the chloroplast illustrated in Fig. 2a,b. The parameters of the calculation correspond to the case a) of Table 1. $D=1.7 \cdot 10^{-5}$ sm$^2$s$^{-1}$ -diffusion coefficient of $CO_2$ in the chloroplast.

The plot of the $CO_2$ concentration distribution function in the chloroplast due to light driven $CO_2$ assimilation is shown in Fig.2a as a function of the space coordinates. The main diameter of the chloroplast has been set to 4 μm and the small diameter of the chloroplast has been set to 2 μm. For the $CO_2$ concentration, eqn. (31) is solved in the grid that is enclosed by the geometrical section of the elliptical form of the chloroplast (surface X-Y in Fig 2b, surface Z-X in Fig 2c, surface Z-Y in Fig 2d). The largest concentrations are seen at the boundary of the chloroplast and also the photosynthetic rate of RuBP production. The vertical profiles of the $CO_2$ concentration presented by Figs.2 a,c,d are found to decrease toward the center of chloroplasts by the chemical reaction with the high assimilation rate. The thickness of the boundary layer can be estimated by the value of 1 μm in Fig.2b. The effect of $CO_2$ concentration distribution on the boundary layer of the chloroplast is consistent with the characteristic time for effective diffusion and the constant of the $CO_2$ assimilation rate. If the light irradiance intensity were artificially increased, the gradient of the $CO_2$ concentration would be enhanced but the thickness of the boundary layer decreased. Note the boundary value of $CO_2$ concentration also plays an important role as governed by the $CO_2$ assimilation rate and the transfer rate or effective diffusion. If the boundary $CO_2$ concentration were decreased, the gradient of the $CO_2$ concentration in the boundary layer of the chloroplast should be decreased as it is represented by Figs. 3 a,b,c.

Figs. 3 a,b,c illustrate the $CO_2$ assimilation flux distribution $A_J$ in the chloroplast volume (surface Z-X in Fig.3 b, surface Z-Y in Fig.3 c).
They coincide with the concentration profile graphs, the high photosynthetic rate occupying the boundary layer of the space of the chloroplast, where the $CO_2$ concentration is elevated from the boundary to the center of the chloroplast. The $CO_2$ flux distribution has the maximum at the surface of the chloroplast and decreases toward the center. Similar to Figs.2 and 3, Figs 4 a,b show the case of low $CO_2$ concentration and assimilation flux in the volume of the chloroplast. The parameters of the calculation appropriate to the case b of Table 1. The calculations were made under conditions of saturating light and low concentration of $CO_2$ in Table 1. From Figs 4 a,b, it can be seen that the gradient of $CO_2$ concentration is not strongly affected by the light irradiation intensity, and the assimilation flux is less by 3 orders of magnitude compared to the case presented in Fig. 3a. Indeed, the boundary condition value of $CO_2$ density under appropriate low sub-stomatal $CO_2$ concentration in mesophyll is a particular feature of these results in that the sink strength distribution in the chloroplast slightly depends on light intensity. Although it is well known to be theoretically possible for $CO_2$ sink strength values to be quite high this depends on the $CO_2$ rate of assimilation and the gradient of concentration to be large and variables of intermediaries of the Calvin cycle change according to the environment. It is believed that this nonuniform distribution of $CO_2$ at the chloroplast boundary layer has been calculated by steady state values of the intermediaries of the reduced Calvin cycle and diffusion of $CO_2$ involved in the more complicated model of the Calvin cycle.

Comparison of Figs. 3 and 4 show that in both the model system that $A_J$ can take large and small values, but the two coefficients take the same sign. Further comparisons show a reasonable quantitative agreement between experimentally observed and model values of $A_J$. In common with $CO_2$ sub-stomatal cavity diffusion the distribution of the calculated $CO_2$ sink strength in the chloroplast volume were made by eqn. (33) and presented in Fig.5. The model response of the photosynthetic $CO_2$ assimilation rate A at the single chloroplast level and that of the $CO_2$ assimilation flux $A_c$ to $I$ are shown in Figs. 5 and 6 respectively. While a photosynthetic $CO_2$ sink strength for one chloroplast is presented by Fig.5, the $CO_2$ assimilation rate calculated at the leaf level is presented in Fig. 6. Taking into account the osmotic volume of chloroplasts as: 25 μl/mg chlorophyll and chlorophyll density 0.5 g per $m^2$ of leaf area, then we get for the $CO_2$ flux density by leaf area: $A_c=A*\delta$ where: $\delta=1.25*10^{-5}$ m. The $CO_2$ sink strength can be evaluated in the case of parameters of a turnover cycle presented by Table 1; case a varies from 0.4 to 0.08 (mol $m^{-3}s^{-1}$).

The curve of the $CO_2$ flux density by leaf area has practical usability to determine the rate constants of light driven RuBP regeneration $k_Y$. If the measurement can be made of $I$ and $A$, then $k_Y$ can be determined (or at least approximated form parameterized data) as the first derivation of the function A(I). The curve of A(I) and eqns. (17), (18), and (32) allow a convenient method for determining $k_I$ and $k_Y$ values; in the model the next measurable parameters and intermediary concentrations are presented by Table 1. As far as values of $A$ are concerned, the quantitative agreement between experiment and the model is good (Poolman, 2000).

## 5. Conclusions

The steady-state expressions for photosynthetic $CO_2$ sink strength considering a two-substance model and incident irradiation driven production rate of high energy compounds coupled with $CO_2$ transport at the chloroplast level are derived. The two substances of light activation compounds and two intermediates of the $CO_2$ assimilation cycle are included in the chain of processes for the regeneration of RuBP.

This approach includes simplifications but is physically consistent with the $CO_2$ diffusion process in chloroplasts and can be utilized when interpreting experimental data on $CO_2$ assimilation rate. The dependence of $CO_2$ concentration and flux on the boundary layer and on the surface of the chloroplast on the light flux was a realistic estimation at the high level of incident irradiation.

It is shown that effective $CO_2$ concentration gradient in the single chloroplast arises from high concentrations of intermediates of the Calvin cycle and metabolites of the light reactions under high level of the incident irradiation. The results of the modelling also suggest that there are high concentrations of Calvin cycle metabolites in the stroma of the chloroplast. It does indeed occur, as was demonstrated previously (Kaitala, 1982), at high carbon assimilation rate.

Although the presented model of leaf photosynthesis is significantly simplified it does not mask the role of light and dark reactions in the stroma of the chloroplast and it can be used for the realization of more complicated environmental models. The model can be utilized in the interpretation of $CO_2$ response measurements in irradiation environment and sucrose production.

The theoretical evidence presented here suggests that light irradiance exerts considerable control over carbon assimilation, ATP production in the Calvin cycle and existence of the gradient of the $CO_2$ concentration over the volume of the chloroplast. The $CO_2$ assimilation flux is generally consistent with the modeling results and a comparison suggests that to understand the behavior of the carbon turnover, the influence of the reaction rate generally considered to be part of the Calvin cycle, but known to be present in the chloroplast stroma, must also be taken into account. The theoretical results also suggest that the metabolic pathway described in the model of the Calvin cycle can be employed in constructing the environmental models of carbon assimilation. On the basis of this photosynthesis model, it would be a challenging subject to implement the electron transfer phenomena (Topmanee et al. 2010; Xu et al. 2010; Kim et al. 2010) on the excitation dyanamic process of S $\rightarrow$ S* at the femtosecond level by using the Z-scheme (Ke, 2003) of the light reaction. From a broader perspective, we hope that this model can serve as a useful theoretical foundation for a more complete and quantitative understanding of a wide range of photosynthesis based processes.

## 6. Acknowledgements

We thank A. Lushnikov and A. Kukushkin for useful scientific discussions.

## 7. References

Acock, B., Hand, D. W., Thornley, J.H.M., and Wilson, W.J., 1976. Photosynthesis in stands of green pepper. An application of empirical and mechanistic models to controlled-environment data. Ann. Bot., **40**, 1293-1307.

Cooke, J.R. 1967. Some theoretical considerations in stomatal diffusion: a field theory approach. Acta Biotheoretica. **XVII**, 95-124.

Farquhar, G.D., von Caemmerer, S., Berry, J.A., 1980. A biochemical model of photosynthetic $CO_2$ assimilation in leaves of $C_3$ species. Planta. **149**, 78-90.

Giersch, C., 1986. Oscillatory response of photosynthesis in leaves to environmental perturbations: a mathematical model. Archives of Biochemistry and Biophysics. **245** (1), 263-270.

Giersch, C., 2003. Stationary diffusion gradients associated with photosynthetic carbon flux – a study of compartmental versus diffusion-reaction models. Journal of Theor. Biol. **224**, 385-379.

Hahn, B. D., 1984. A mathematical model of leaf carbon metabolism. Ann. Bot. **54,** 325-339.

Hahn, B. D. 1987. A mathematical model of photorespiration and photosynthesis. Ann. Bot. **60**, 157-169.

Hahn, B. D. 1991. Photosynthesis and photorespiration: modelling the essentials. J. Theor. Biol. **151**, 123-139.

Kaitala, V., Hari, P., Vapaavuori, E. and Salminen, R. 1982. A dynamic model for photosynthesis. Ann. Bot. **50**, 385-396.

Karavaev V.A., Kukushkin A.K. 1993. A theoretical model of light and dark processed of photosynthesis: the problem of regulation. Biophysics **38** (16), 957-975.

Kim, W. Y., Choi, Y. C., Min, S. K., Cho, Y. C., and Kim, K. S. 2009. Application of quantum chemistry in nanotechnology: electron/spin transport in molecular devices. Chem. Soc. Rev. 38, 2319-2333.

Ke Bacon, Photosynthesis: Photobiochemistry and Photobiophysics, Advances in Photosynthesis, Vol.10. 2003. Kluwer Academic Publishers. pp 747.

Kukushkin A.K., Tikhonov A.N. Lectures on the biophysics of plant photosynthesis. Moscow University press, Moscow, 1988, pp 320.

Laisk, A. 1973. Mathematical model of photosynthesis and photorespiration. Reversible phosphoribulokinase reaction. Biophysics **18,** 679-684.

Laisk, A., Eichelmann, H., Oja, V., Eatherall, A., and Walker, D.,A. 1989. A mathematical model of the carbon metabolism in photosynthesis. Difficulties in explaining oscillations by fructose 2,6-bisphosphatase. Proc. R. Soc. London, B, **237**, 417-444.

Lushnikov, A.A., Ahonen, T., Vesala, T., Juurola, E., Nikinmaa, E., and Hari, P. 1997. Modelling of light-driven RuBP regeneration, carboxylation and $CO_2$ diffusion for leaf photosynthesis. J. Teor.Biol. **188.** 143-151.

Lushnikov, A.A., Vesala, T., Kulmala, M., and Hari, P., 1994. A semiphenomenological model for stomatal gas transport. J. Teor. Biol. **171.** 291-301.

Milstein, J. and Bremermann H.J., 1979. Parameter identification of the Calvin photosynthesis cycle. J. Math. Biol. **7**, 99-116.

Massunaga, M.S.O., Gatts, C. E.N., Gomes, A.G., and Vargas, H., 2001. A simple model for dynamics of photosynthesis. Analytical sciences. April, **17**, Special Issue. S29-S30.

Nitzan, A., Ross, J., 1973. Oscillations, multiple steady states, and instabilities in illuminated systems. J. Chem. Phys. **59** (1), 241-250.

Nitzan, A., Chemical dynamics in condensed phases. Relaxation, transfer and reaction in condensed molecular system. 2006. Oxford University Press, New York. pp. 709

Pettersson, G., Ryde-Pettersson, U. 1988. A mathematical model of the Calvin photosynthesis cycle. Eur. J. Biochem. **175**, 661-672.

Poolman, M.,G., Fell, D.,A., Thomas, S., 2000. Modelling photosynthesis and its control. J.Exp.Bot. **51,** 319-328

Sveshnikov, A.,G., Bogolubov, A.,N., Kravzov, V.,V., 1993. Lectures on mathematical physics. Moscow University press, Moscow. pp 325.

Thornley, J.H.M., 1974. Light fluctuation and photosynthesis. Ann. Bot. **38**, 363-373.

Thornley, J.H.M. 1976. Mathematical Models in Plant Physiology. Academic Press, London, and New York.

Tipmanee, V., Oberhofer, H., Park, M., Kim, K. S., Blumberger J., 2010. Prediction of reorganization free energies for biological electron transfer: A comparative study of Ru-modified cytochrome and a 4-helix bundle protein. J. Am. Chem. Soc. 132, 17032-17040.

Ustin, S.L., Jacquemoud, S., and Govaerts, Y., 2001. Simulation of photon transport in a three-dimensional leaf: implications for photosynthesis. Plant, Cell and Environment **24**, 1095-1103.

Xu, Z., Singh, N. J., Pan J., Kim, H., Kim, K. S., and Yoon, J., 2009. A Unique Sandwich Stacking of Pyrene-Adenine-Pyrene for Selective and Ratiometric Fluorescent Sensing of ATP at Physiological pH, J. Am. Chem. Soc. 131, 15528-15533

Table 1. The parameters of the model.

| Definition | Value | |
|---|---|---|
| | $\xi C/S^* >> 1$ (a) | $\xi C/S^* << 1$ (b) |
| $k_I$   (µmol quanta)$^{-1}$m$^2$ | 0.16[a,b] | 0.16[a,b] |
| $k_X$   (mol m$^{-3}$)$^{-1}$s$^{-1}$ | 0.344[c] | 0.344[c] |
| $I$ (µmol quanta) m$^{-2}$ s$^{-1}$ | 1000 | 1000 |
| $\xi = k_X / k_Y$ | 25.3[c] | 1.521[c] |
| $\gamma = S/S^*$ | 0.282[d] | 0.167[d] |
| $X_0$   (mol m$^{-3}$) | 6[c] | 4[c] |
| $S^*$   (mol m$^{-3}$) | 0.39[d] | 0.36[d] |
| $c_0$   (mol m$^{-3}$) | 0.2 | 0.012 |
| A   (mol m$^{-3}$ s$^{-1}$) | 0.4 | 0.3 |
| A$_c$   (µmol) m$^{-2}$s$^{-1}$ | 8.1 | 1.3 |

[a]Kaitala, 1982;
[b]Lushnikov, et al. 1997;
[c]Hahn, 1984, 1987, 1991;
[d]Pettersson, et al. 1988.

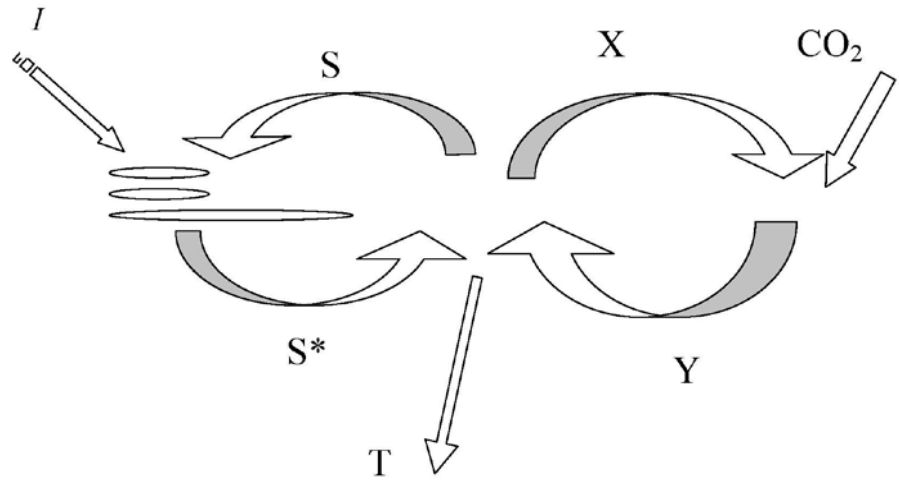

Fig.1. Scheme represents a two substance model for the photosynthesis at the single chloroplast level. The light reactions of metabolites of $S^*$ (ATP) and $S$ (ADP) are activated by light irradiance with flux $I$. The dark reactions of the Calvin cycle are presented by the turnover of two pairs of compounds X (RuBP) and Y (PGA). The triose (T) denotes the output from the Calvin cycle.

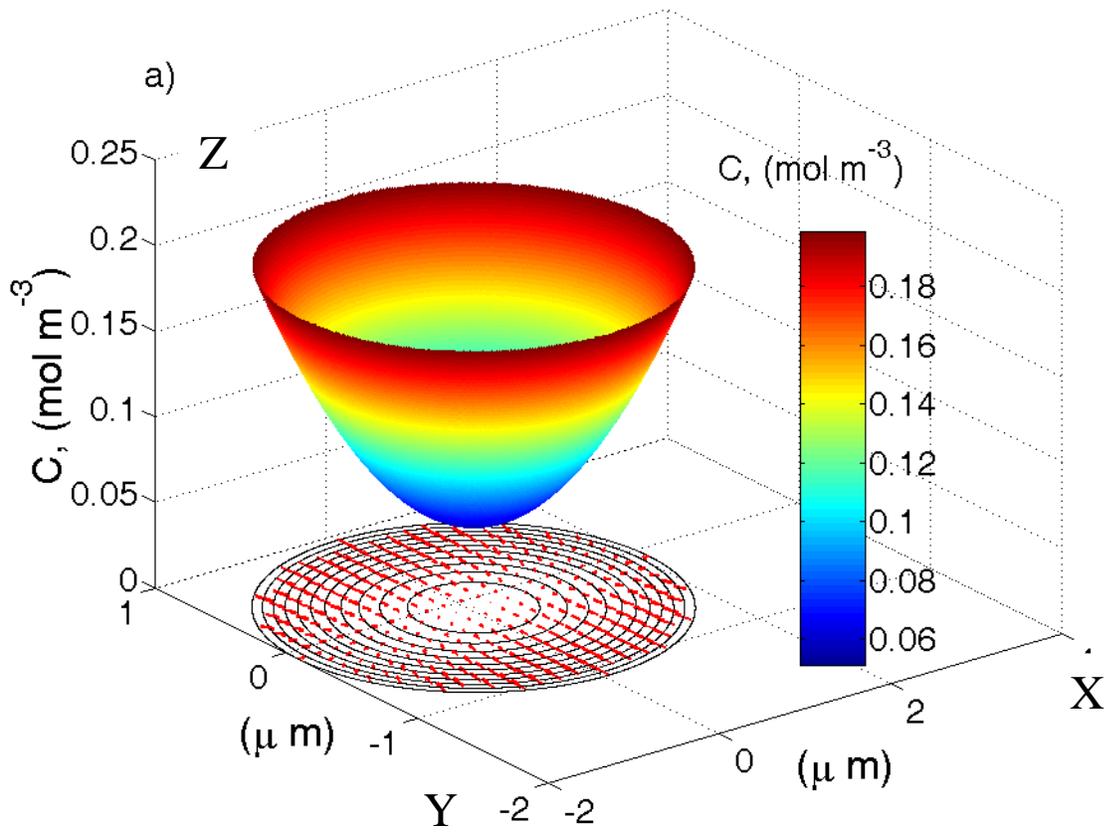

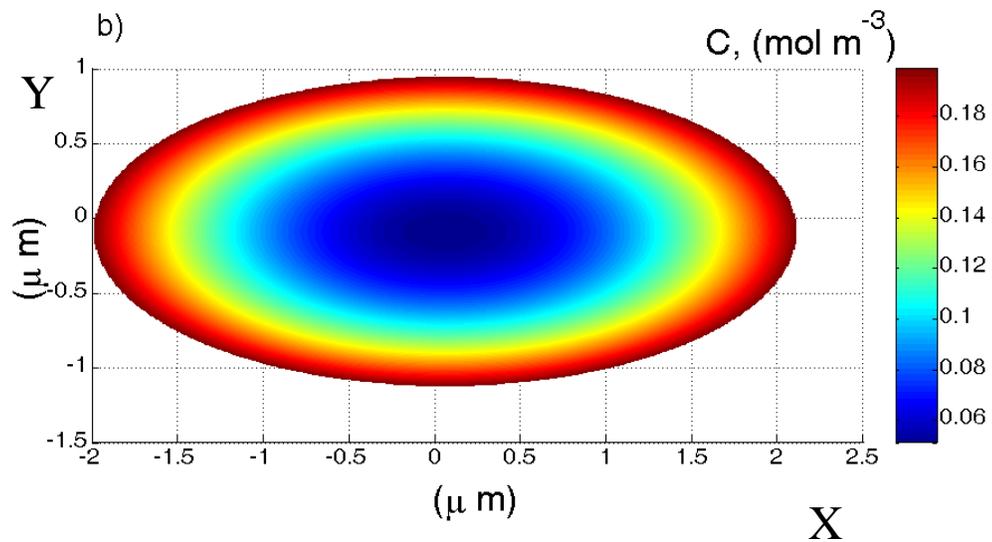

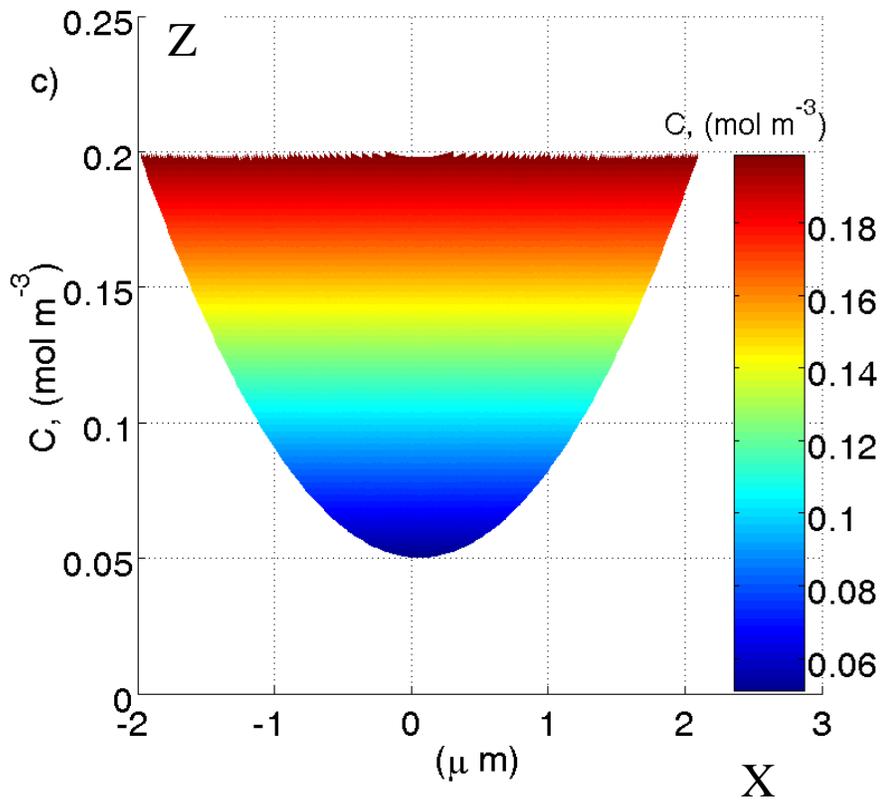

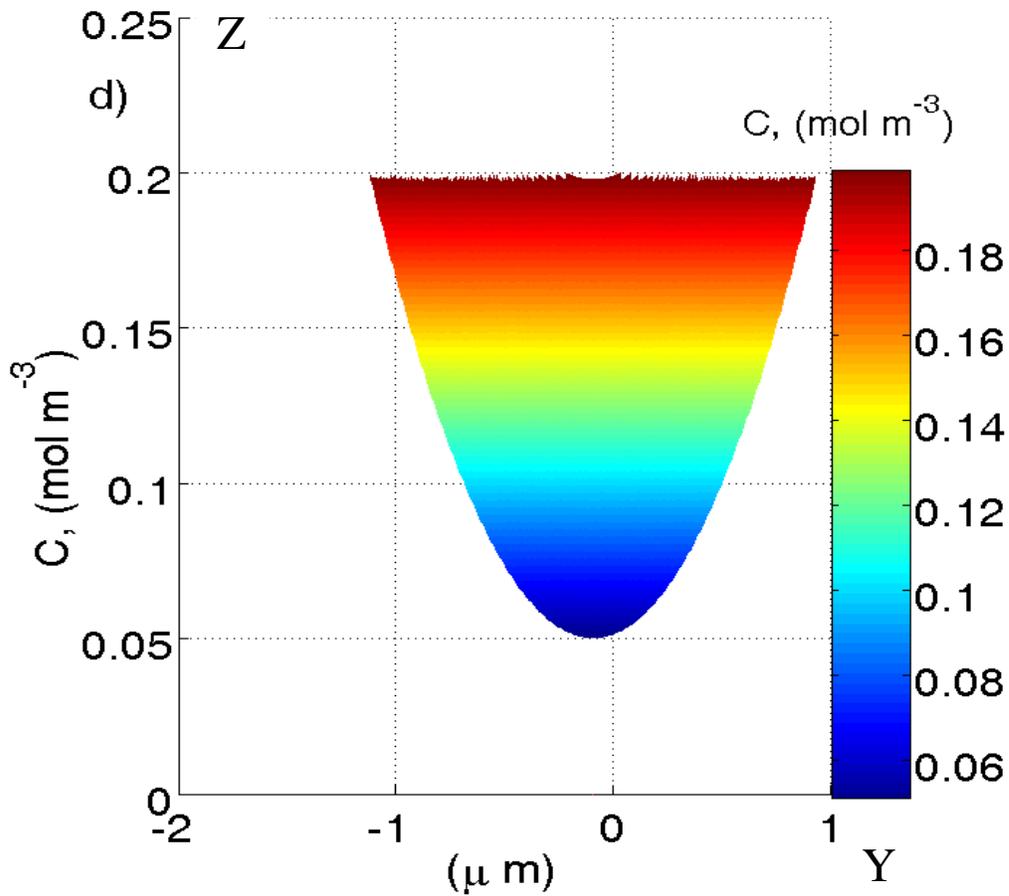

Fig. 2. a) $CO_2$ concentration profile in the chloroplast (max concentration, c: 0.2 mol/m$^3$) , b) cross section of the ellipsoidal chloroplast on the x-y plane [$x^2 + (y/2)^2 = 1$; $|x| \leq 1$ μm, $|y| \leq 2$ μm]. c) $CO_2$ concentration profile on the x-z cross section, d) $CO_2$ concentration profile on the y-z cross section.

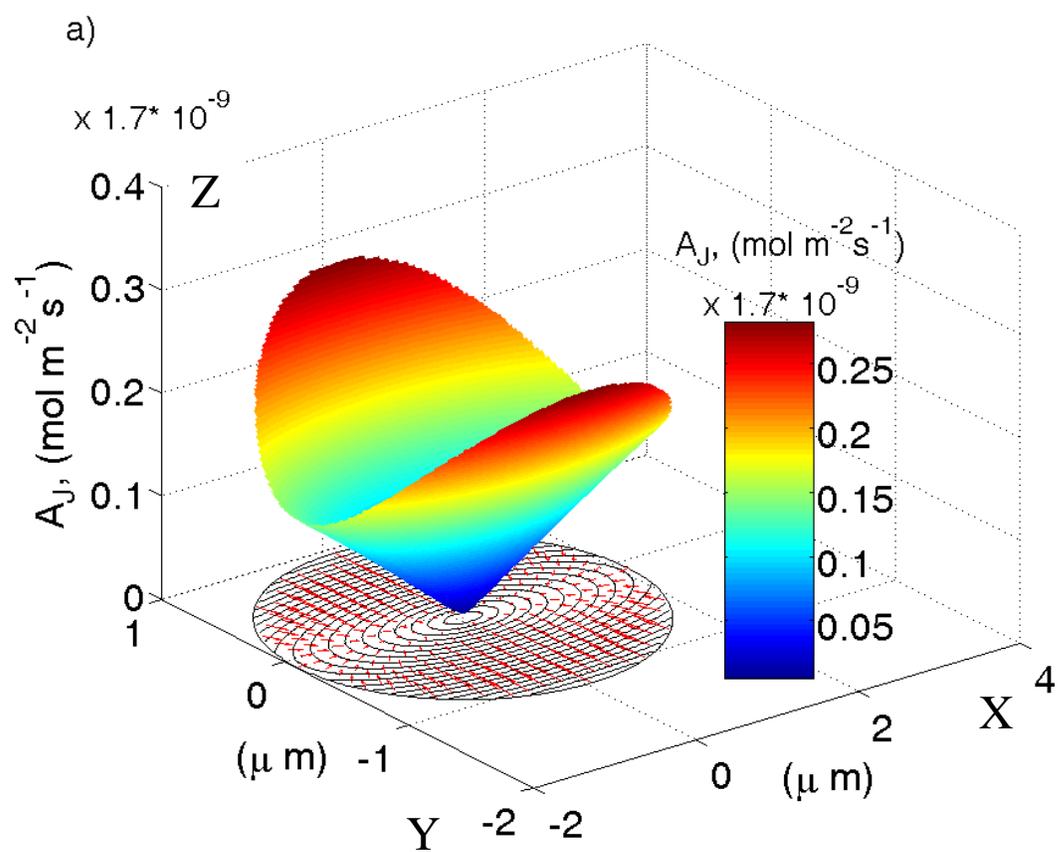

a)

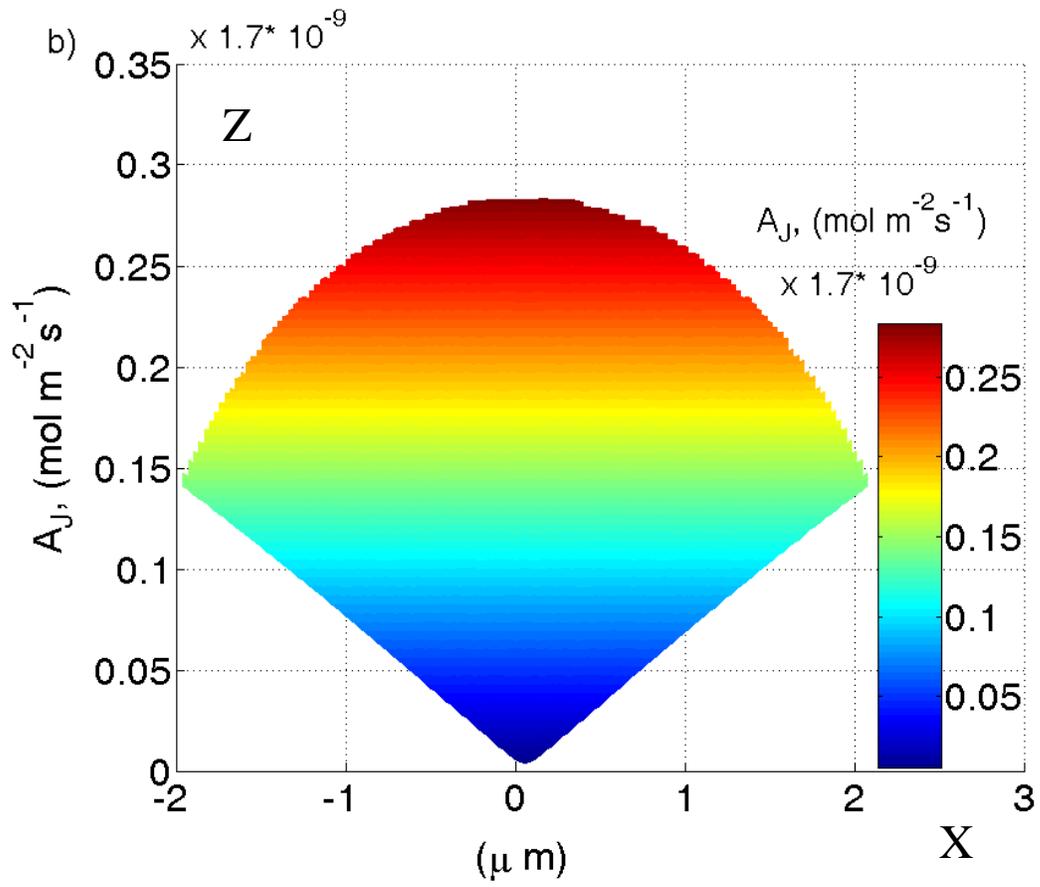

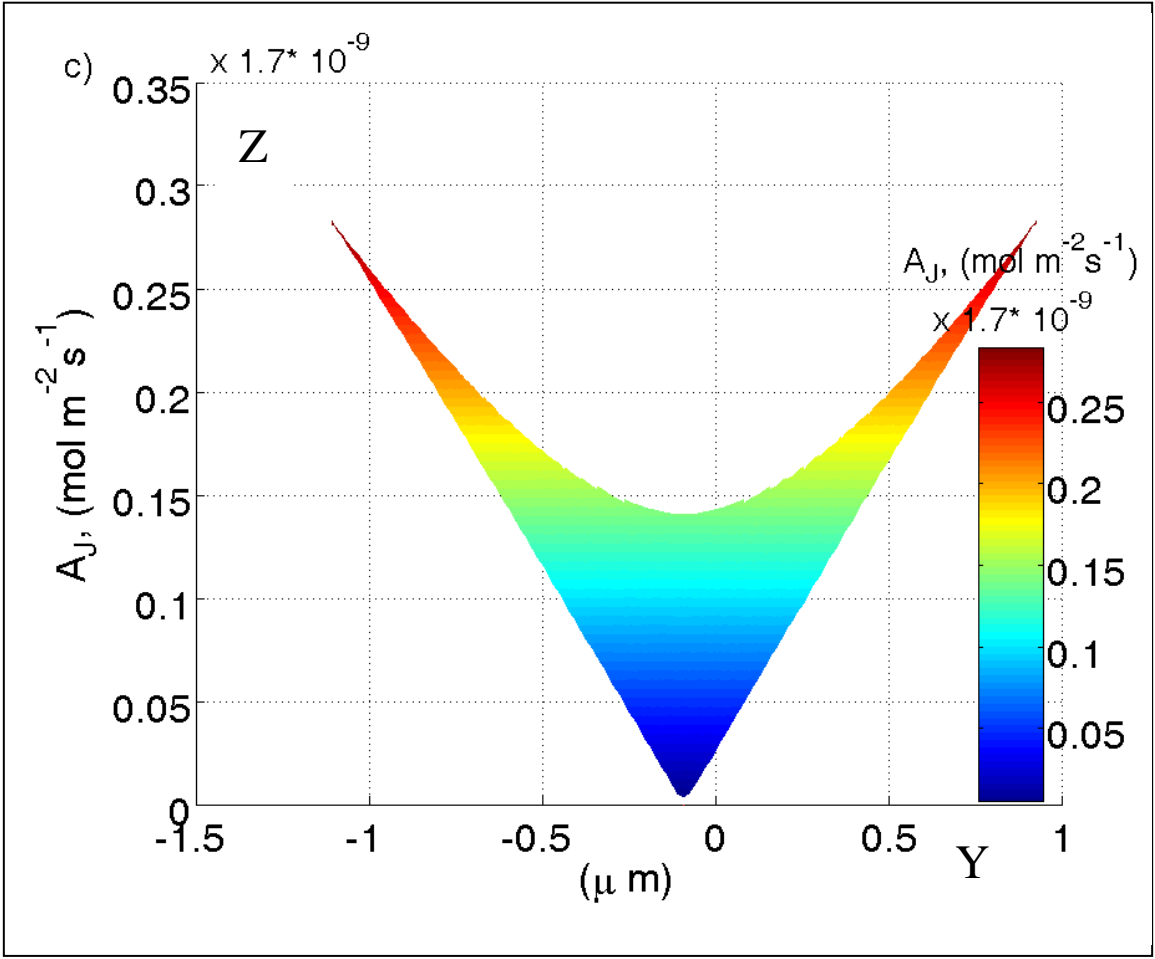

Fig.3. a) profile of the $CO_2$ assimilation flux at the single chloroplast level, b) profile of the $CO_2$ assimilation flux on the x-z cross section, c) profile of the $CO_2$ assimilation flux on the y-z cross section

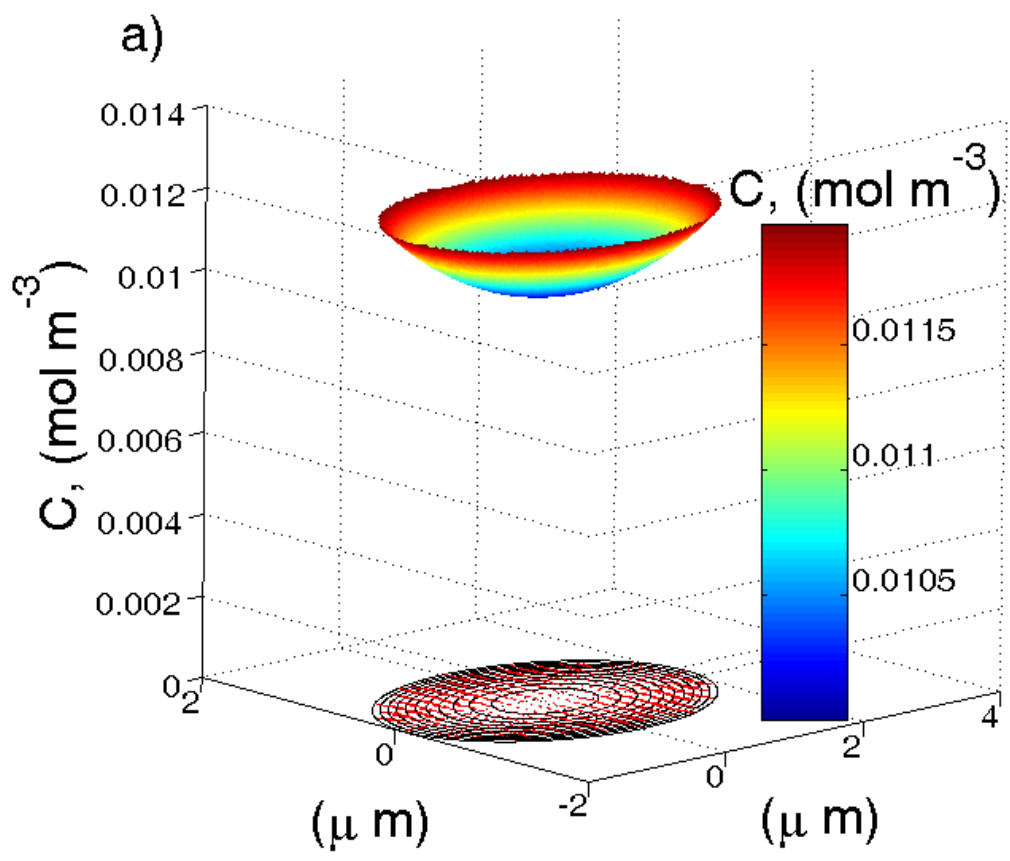

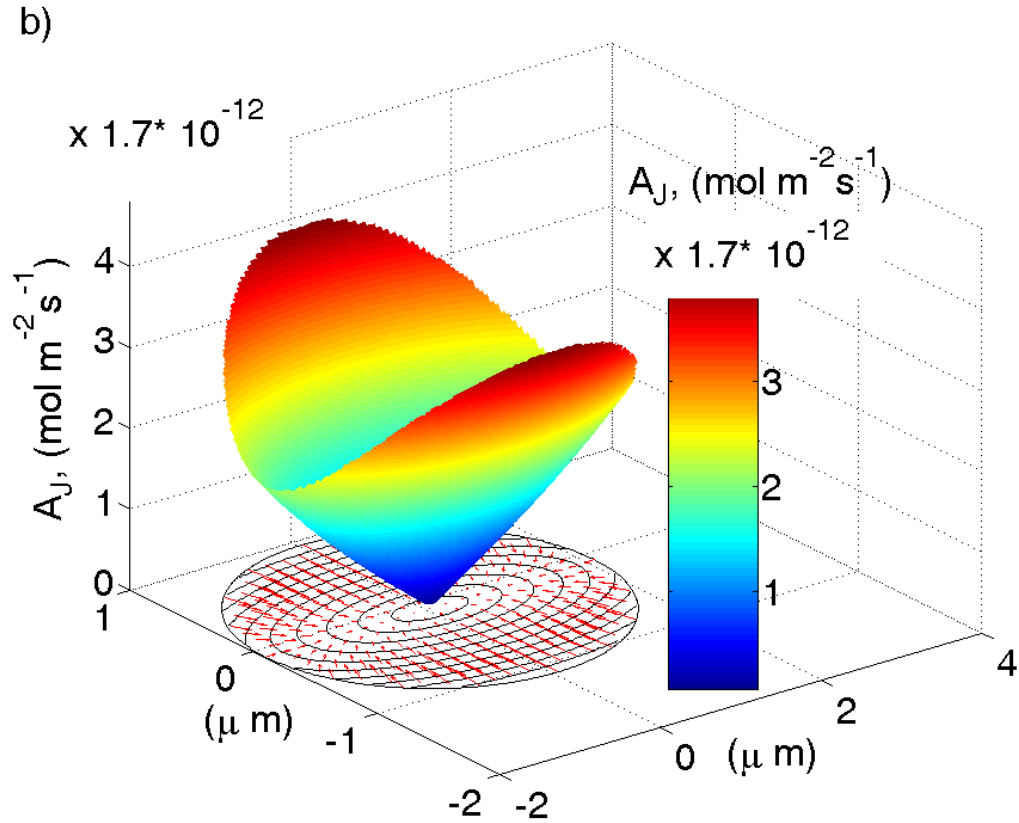

Fig. 4. a) $CO_2$ concentration profile in the chloroplast under low $CO_2$ boundary concentration $c_0 = 0.012 \, (mol \, m^{-3})$, b) $CO_2$ assimilation flux of chloroplast under low $CO_2$ boundary concentration $c_0 = 0.012 \, (mol \, m^{-3})$.

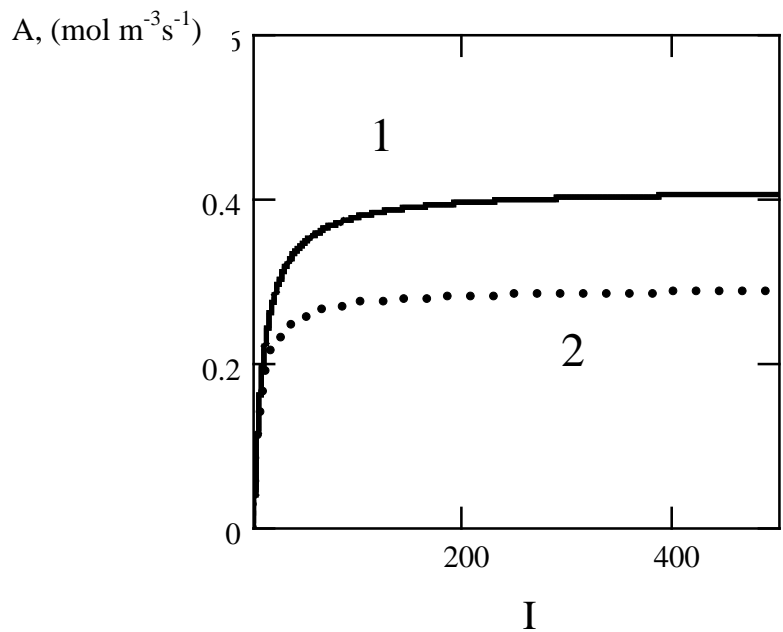

Fig. 5. $CO_2$ sink strength at the single chloroplast level as a function of photon flux density. Curve (1) the boundary value of $CO_2$ concentration saturation point is $c_0 = 0.2$ (mol/m$^3$), curve (2) the boundary value of $CO_2$ concentration is $c_0 = 0.14$ (mol/m$^3$).

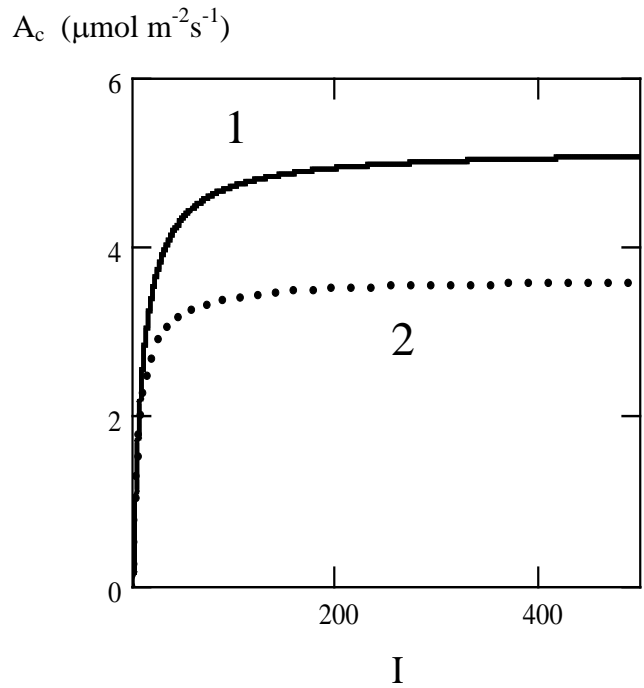

Fig. 6. Calculated $CO_2$ assimilation flux density ($A_c$) at the leaf level as a function of photon flux density. Curve (1) the boundary value of $CO_2$ concentration saturation point is $c_0 = 0.2$ (mol/m$^3$), and curve (2) the boundary value of $CO_2$ concentration is $c_0 = 0.14$ (mol/m$^3$).